# Label-free evaluation of lung and heart transplant biopsies using tissue autofluorescence-based virtual staining

Yuzhu Li[†,1,2,3], Nir Pillar[†,1,2,3], Tairan Liu[1,2,3], Guangdong Ma[1], Yuxuan Qi[4], Kevin de Haan[1,2,3], Yijie Zhang[1,2,3], Xilin Yang[1,2,3], Adrian J. Correa[5], Guangqian Xiao[5], Kuang-Yu Jen[6], Kenneth A. Iczkowski[6], Yulun Wu[7], William Dean Wallace[5], and Aydogan Ozcan[*,1,2,3,8]

[1]Electrical and Computer Engineering Department, University of California, Los Angeles, CA, 90095, USA.

[2]Bioengineering Department, University of California, Los Angeles, CA, 90095, USA.

[3]California NanoSystems Institute (CNSI), University of California, Los Angeles, CA, 90095, USA.

[4]Department of Computer Science, University of California, Los Angeles, 90095, USA.

[5]Department of Pathology, Keck School of Medicine, University of Southern California, Los Angeles, CA, 90033, USA.

[6]Department of Pathology and Laboratory Medicine, School of Medicine, University of California, Davis, Sacramento, CA, 95817, USA.

[7]Department of Mathematics, University of California, Los Angeles, CA, 90095, USA.

[8]Department of Surgery, University of California, Los Angeles, CA, 90095, USA.

*Correspondence: Aydogan Ozcan, ozcan@ucla.edu

[†] Equal contributing authors

## Abstract

**Objective and Impact Statement:** We present a panel of virtual staining neural networks for lung and heart transplant biopsies, providing rapid and high-quality histological staining results while bypassing the traditional histochemical staining process. **Introduction:** Allograft rejection is a common complication of organ transplantation, which can lead to life-threatening outcomes if not promptly managed. Histological examination is the gold standard method for evaluating organ transplant rejection status, as it provides detailed insights into rejection signatures at the cellular level. Nevertheless, the traditional histochemical staining process is time-consuming, costly, and labor-intensive since transplant biopsy evaluations typically necessitate multiple stains. Furthermore, once these tissue slides are stained, they cannot be reused for other ancillary tests. More importantly, suboptimal handling of very small tissue fragments from transplant biopsies may impede their effective histochemical staining, and color variations across different laboratories or batches can hinder efficient histological analysis by pathologists. **Methods:** To mitigate these challenges, we developed a panel of virtual staining neural networks for lung and heart transplant biopsies, which digitally convert autofluorescence microscopic images of label-free tissue sections into their brightfield histologically stained counterparts – bypassing the traditional histochemical staining process. Specifically, we virtually generated Hematoxylin and Eosin (H&E), Masson's Trichrome (MT), and Elastic Verhoeff-Van Gieson (EVG) stains for label-free transplant lung tissue, along with H&E and MT stains for label-free transplant heart tissue. **Results:** Blind evaluations conducted by three board-certified pathologists confirmed that the virtual staining networks consistently produce high-quality histology images with high color uniformity, closely resembling their well-stained histochemical counterparts across various tissue features. The use of virtually stained images for the evaluation of transplant biopsies achieved comparable diagnostic outcomes to those obtained via traditional histochemical staining, with a concordance rate of 82.4% for lung samples and 91.7% for heart samples. Moreover, virtual staining models create multiple stains from the same autofluorescence input, eliminating structural mismatches observed between adjacent sections stained in the traditional workflow, while also saving tissue, expert time, and staining costs. **Conclusion:** The presented virtual staining panels provide an effective alternative to conventional histochemical staining for transplant biopsy evaluation. These virtual staining panels have the potential to enhance the clinical diagnostic workflow for organ transplant rejection and improve the performance of downstream automated models for the analysis of transplant biopsies.

## Introduction

Transplantation is the only mode of therapy for most cases of end-stage organ failure. Lung transplantation serves as the ultimate therapeutic option for many patients with end-stage lung disease. Similarly, cardiac transplantation is the treatment of choice for many patients with end-stage heart failure who remain symptomatic despite optimal medical therapy. According to the latest report from the International Society for Heart and Lung Transplantation (ISHLT), 29% of adult lung transplant patients experience at least one episode of treated acute rejection between hospital discharge and 1-year follow-up after transplant[1]. Nearly a quarter of cardiac transplanted patients undergo treated rejection within 1 year[2]. Although several clinical monitoring methodologies utilize noninvasive modalities to identify transplant-related complications, such as pulmonary function tests, laboratory analyses (including microbiological or immunological tests), and genomic tests (such as donor-derived cell-free DNA[3]), the diagnosis of allograft rejection continues to rely on the histological assessment of organ biopsies, which is also important to exclude other diagnostic entities such as infections or drug toxicity. Specifically, acute lung rejection is diagnosed when lung tissue shows perivascular and interstitial mononuclear cell infiltration[4]. Pathologists typically evaluate these tissue slides using the hematoxylin and eosin (H&E) stain, along with additional special stains that highlight relevant areas on the slide. For instance, connective tissue is highlighted by the Masson Trichrome (MT) stain, while elastic fibers are emphasized using the elastic Verhoeff-Van Gieson (EVG) stain. Similarly, cardiac acute cellular rejection is diagnosed based on the identification of interstitial or perivascular infiltrates, composed of lymphocytes, variable macrophages, and rarely eosinophils, with or without foci of myocyte damage[5]. To evaluate cardiac transplant biopsies, pathologists examine H&E slides, which provide a broader view of tissue architecture and inflammation, along with MT to evaluate the level of fibrosis.

Although traditional histochemical staining serves as the gold standard for evaluating transplant rejection cases, it still faces several challenges. First, it necessitates a specialized laboratory infrastructure and the expertise of skilled histotechnologists to perform laborious tissue preparation and staining procedures for multiple stains (including several special stains), which is time-consuming, reagent-intensive and costly. Since every tissue section is stained with a distinct stain, differences naturally occur between adjacent sections, and this leads to feature mismatches. Pathologists must toggle between different stain views and manually align relevant regions, a process prone to errors and inefficiencies. Secondly, the tissue slides after histochemical staining cannot be preserved or reused for other supplementary tests. Moreover, the small tissue fragments extracted by transbronchial biopsies or catheter biopsies of transplant recipients are susceptible to mishandling and errors in the histochemical staining process, resulting in various levels of staining artifacts. Specifically, the tiny lung specimens, composed of small portions of alveolar tissue, are easily crushed and distorted during the biopsy procedure, leading to uneven tissue fixation, improper embedding, and uneven staining or overstaining of tissue[6]. These morphological changes and artifacts render histological interpretation difficult to draw deterministic conclusions and significantly reduce the degree of agreement among diagnosticians[7]. Similarly, procedure-related artifacts are also frequently encountered in cardiac biopsies and can result in interpretive difficulties, including tissue compression, acute hemorrhage related to the procedure, and artifactual contraction-band change. Furthermore, staining variations across different laboratories or batches can impede efficient histological analysis, posing another challenge for pathologists in achieving consistent diagnostic results.

Recently, the rapid advances in deep learning technologies have introduced transformative opportunities to a wide range of biomedical research problems and clinical diagnostics[8–10]. For example, one notable application of these technologies is the use of deep learning for virtual histological staining of label-free tissue samples[11–15]. In this technique, a deep neural network (DNN) is trained to computationally stain microscopic images of label-free tissue sections, matching their histologically stained counterparts. Deep learning-enabled label-free virtual staining aims to circumvent some of the challenges associated with the traditional histochemical staining workflow, such as time consumption, labor intensity, high costs, stain variations among labs/technicians, and staining-related artifacts. Multiple research groups explored deep learning-based virtual staining techniques and successfully applied this technology to generate a range of routinely used stains, both histological[16–39] and immunohistochemical[40–42].

However, these earlier studies have primarily focused on core needle biopsies, which output a 1-2 cm long cylinder-shaped tissue[43,44]. This is considerably larger than the 1-3 mm fragments typically obtained for heart/lung transplant biopsies. As a result, they are less susceptible to the crushing and distortion artifacts that frequently afflict transbronchial and endomyocardial biopsies composed of tiny tissue fragments. Addressing this gap, in our study, we demonstrate a panel of virtual tissue staining DNN models for label-free lung and heart transplant biopsy samples, which can consistently and rapidly provide high-quality histologically stained images for the evaluation of acute transplant rejection. As depicted in Fig. 1, we employed five separate DNN models to digitally transform autofluorescence microscopic images (including DAPI, TxRed, FITC and Cy5 channels of autofluorescence) of label-free lung and heart transplant biopsies into their corresponding brightfield versions, virtually creating H&E (heart and lung), MT (heart and lung) and EVG (lung only) stains. By utilizing the same autofluorescence input, virtually stained images of different types of stains were generated, producing perfectly registered tissue images at the nanoscale. This approach effectively resolves feature mismatches caused by tissue heterogeneity between adjacent tissue cuts – a common issue in traditional histochemical staining workflows. These virtually stained histology images demonstrate a strong match to their well-stained histochemical counterparts in the appearances of cellular details, such as nuclear shape, cytoplasm, and extracellular features across the tiny tissue fragments. This high degree of similarity was blindly confirmed through score-based quantitative evaluations conducted by three board-certified pathologists. Furthermore, the diagnostic concordance among human diagnosticians who blindly classified transplant biopsies, stained with either virtual or histochemical staining, was high, with 82.4% agreement for lung samples and 91.7% agreement for heart samples. Our results and analyses will pave the way for further large-scale applications of virtual tissue staining in transplant biopsy evaluations. In addition, the consistent and rapid outputs from virtual tissue staining have the potential to benefit the development of downstream algorithms for automatically evaluating transplant biopsies, further improving the efficiency and consistency of organ transplant diagnostics.

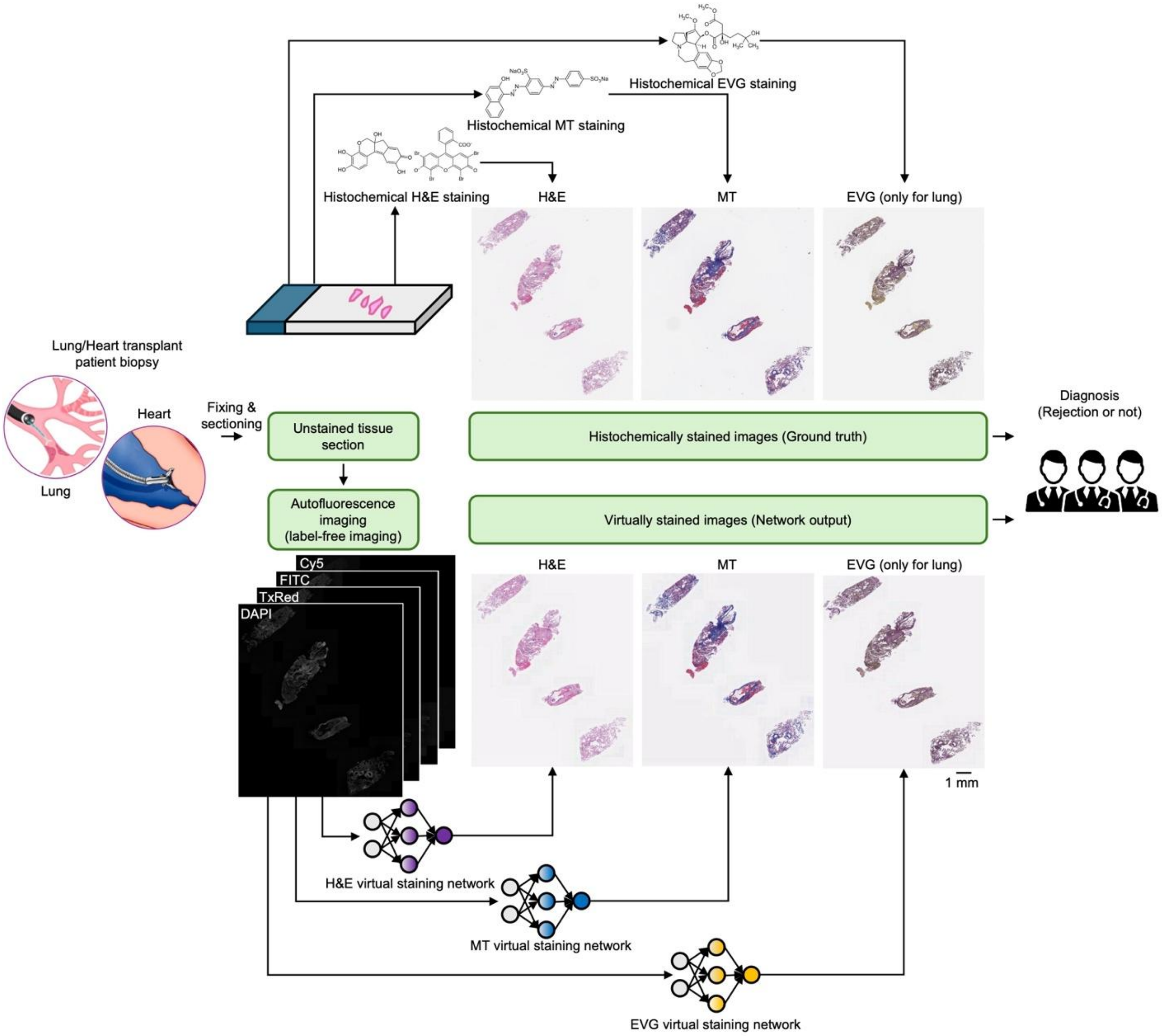

**Figure 1. Deep learning-enabled virtual staining panels for evaluating lung and heart transplant biopsies.** Compared to traditional histochemical staining (top), virtual tissue staining (bottom) can digitally transform autofluorescence images of label-free transplant tissue sections into their histochemically stained equivalents (H&E, MT, and EVG for lung; H&E and MT for heart) with consistent high-quality. Utilizing these virtually stained WSIs, pathologists can achieve diagnostic results

comparable to those obtained with the standard histochemical staining workflow, achieving a shorter turnaround time while also preserving tissue for further analysis and saving costs.

# Results

### Virtual staining of unlabeled tissue sections from lung and heart transplant patients

To facilitate label-free post-operative evaluation of lung and heart transplant biopsies and overcome the challenges of traditional histochemical staining (e.g., stain variations, lengthy turnaround time, exhaustion/depletion of tissue blocks for multiple chemical stains, etc.), we developed a suite of lung and heart virtual staining network models (5 models in total) to digitally transform autofluorescence images of unlabeled lung/heart tissue sections into their histochemical equivalents, virtually creating H&E, MT, and EVG stains for lung tissue as well as H&E and MT stains for heart tissue. Each virtual staining DNN model was separately trained based on the structurally-conditioned generative adversarial network (GAN) framework[45,46] as shown in Supplementary Figure 1. The training dataset included a total of 5958 image patches (3303 for lung and 2655 for heart), where each image had 2048×2048 pixels (~333×333 μm²), all sourced from 36 distinct patients – 18 for lung and 18 for heart. The overall training data encompassed ~258 GB, comprising ~23.5 billion unique pixels, i.e., ~23.5 billion histochemically stained ground truth labels, each approximately at the diffraction limit of light.

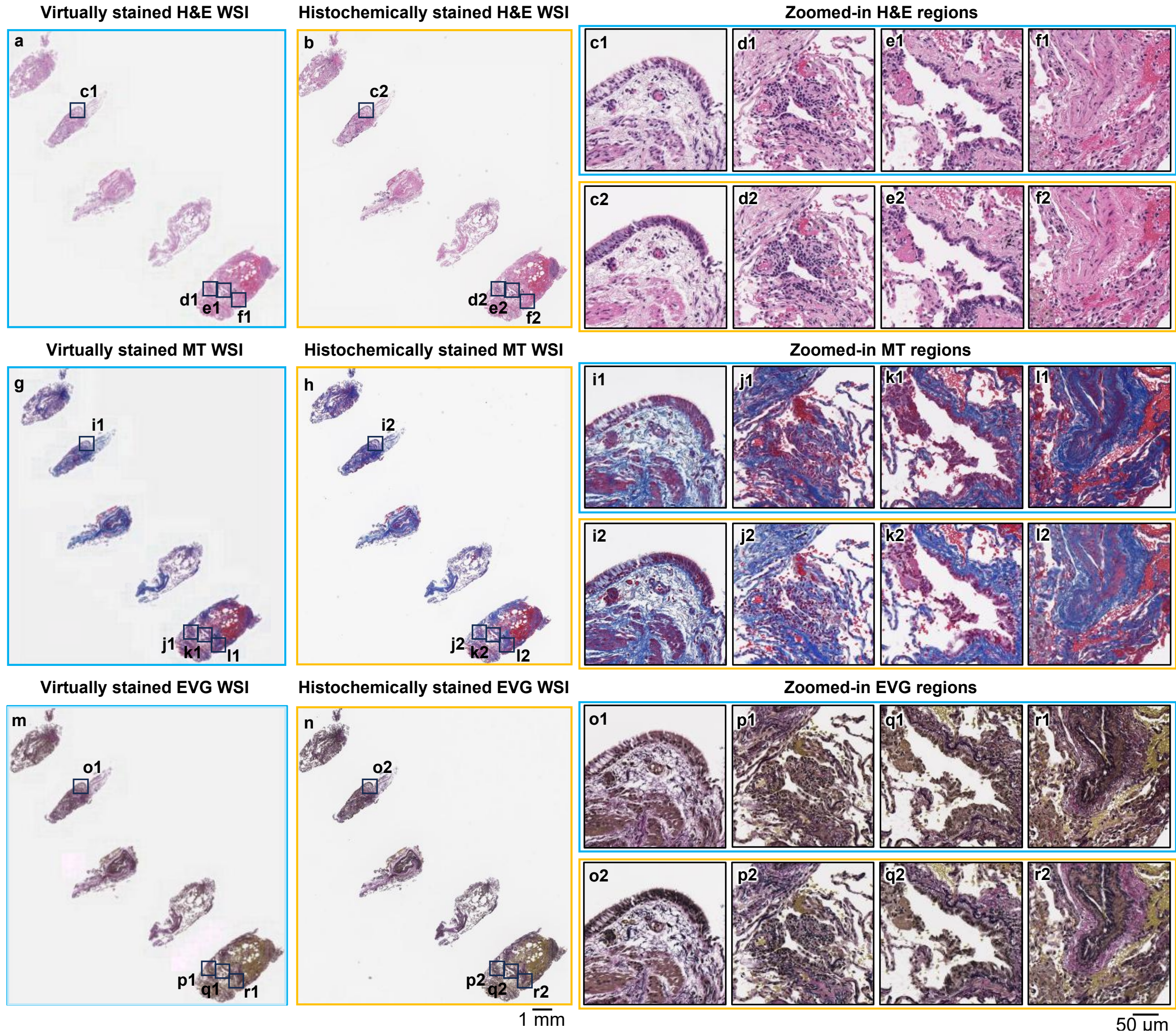

**Figure 2. Visual comparisons between the virtually stained H&E, MT, and EVG images of a lung transplant rejection sample and their corresponding histochemically stained counterparts.** (a) Virtually stained H&E WSI, which was digitally generated by our virtual staining lung-H&E network by taking label-free autofluorescence images as its input. (b) Histochemical H&E staining results of the same WSI in (a), serving as the ground truth. (c-f) Zoomed-in images of the four exemplary local regions indicated in (a-b), where (c1-f1) are the virtually stained H&E images and (c2-f2) are the corresponding histochemically stained H&E images. (g) Virtually stained MT WSI, which was digitally generated by our virtual staining lung-MT network by taking label-free autofluorescence images as its input. (h) Histochemical MT staining results of the same WSI in (g), serving as the ground truth. (i-l) Zoomed-in images of the four exemplary local regions indicated in (g-h), where (i1-l1) are the virtually stained MT images and (i2-l2) are the corresponding histochemically stained MT images. (m) Virtually stained EVG WSI, which was digitally generated by our virtual staining lung-EVG network by taking label-free autofluorescence images as its input. (n) Histochemical EVG staining results of the same WSI in (m), serving as the ground truth. (o-r) Zoomed-in images of the four exemplary local regions indicated in (m-n), where (o1-r1) are the virtually stained EVG images and (o2-r2) are the corresponding histochemically stained EVG images.

After the virtual staining lung DNN models were trained, virtually stained H&E, MT, and EVG whole slide images (WSIs) were generated for 31 unique patients as the blind testing set, with each WSI corresponding to one of the three adjacent unlabeled tissue slides per patient. The overall lung testing data size for all three stain types is ~3.85 Terabytes (TB), including ~33,200 unique fields-of-view (FOVs), each with 3248×3248 pixels. Figure 2 summarizes the comparison of virtual H&E, MT, and EVG images inferred by our virtual staining DNN models against their corresponding bright-field images after the standard histochemical H&E, MT, and EVG staining. The virtually stained images demonstrate a high level of agreement with their histochemically stained counterparts, accurately representing diagnostically relevant features and facilitating effective evaluation of transplant biopsies. For example, a prominent mononuclear cell infiltrate around a venule was observed in the virtually stained H&E image (Fig. 2(d1)), suggesting acute cellular rejection, and matching very well to its histochemically stained counterpart as shown in Fig. 2(d2). Moreover, the circumferential lymphocytic infiltrate observed around the bronchiole in the virtually stained H&E image (Fig. 2(e1)) indicates bronchial inflammation, which closely aligns with the histochemical H&E staining result shown in Fig. 2(e2). In addition, collagen distribution in the virtually stained MT images (Fig. 2(i1)-(l1)) matched accurately with their histochemically stained equivalents (Fig. 2(i2)-(l2)). A similar resemblance between the virtually and histochemically stained images was seen in the EVG stained images, where the elastic fibers demonstrated significant overlap between virtually (Fig. 2(o1)-(r1)) and histochemically stained images (Fig. 2(o2)-(r2)). The accordance between the virtual and histochemical staining results underscores the potential clinical utility of our deep learning-based virtual staining technique for evaluating the rejection status in lung transplants. Higher-magnification insets from Fig. 2 are presented in Supplementary Figures 2-4 for H&E, MT, and EVG, respectively. Supplementary Figure 5 further validates the agreement between the virtual and histochemical staining results, where we evaluated our neural network models on a non-rejection case as a control comparison, further supporting our conclusions. Similarly, higher-magnification insets from Supplementary Figure 5 are shown in Supplementary Figures 6–8 for H&E, MT, and EVG, respectively.

Similarly, we validated our virtual staining heart neural network models on a total of 33 unique patients previously unseen by the networks, yielding a testing dataset of ~2.39 TB with ~20,600 unique FOVs (each with 3248×3248 pixels). For a heart transplant recipient diagnosed with acute cellular rejection, we virtually generated H&E and MT stained WSIs using our virtual staining DNN models and compared those to their histochemically stained counterparts. The visualization results are shown in Fig. 3. The inflammatory infiltrates, observed in virtually stained H&E images (Fig. 3(c1)-(f1)) are consistent with those presented by their corresponding histochemically stained images (Fig. 3(c2) -(f2)), corroborating the diagnosis of heart transplant rejection status for this case. Moreover, the collagen distributions in virtually stained MT images (Fig. 3(i1)-(l1)) are in high concordance with those observed in their histochemically stained equivalents (Fig. 3(i2)-(l2)). Higher-magnification insets from Fig. 3 are provided in Supplementary Figures 9 and 10 for H&E and MT, respectively. Further validating the performance of the heart virtual staining DNN models, an additional case, pertaining to a non-rejection patient, is also demonstrated in Supplementary Figure 11. Higher-magnification insets from that control case (Supplementary Figure 11) appear in Supplementary Figures 12 (H&E) and 13 (MT).

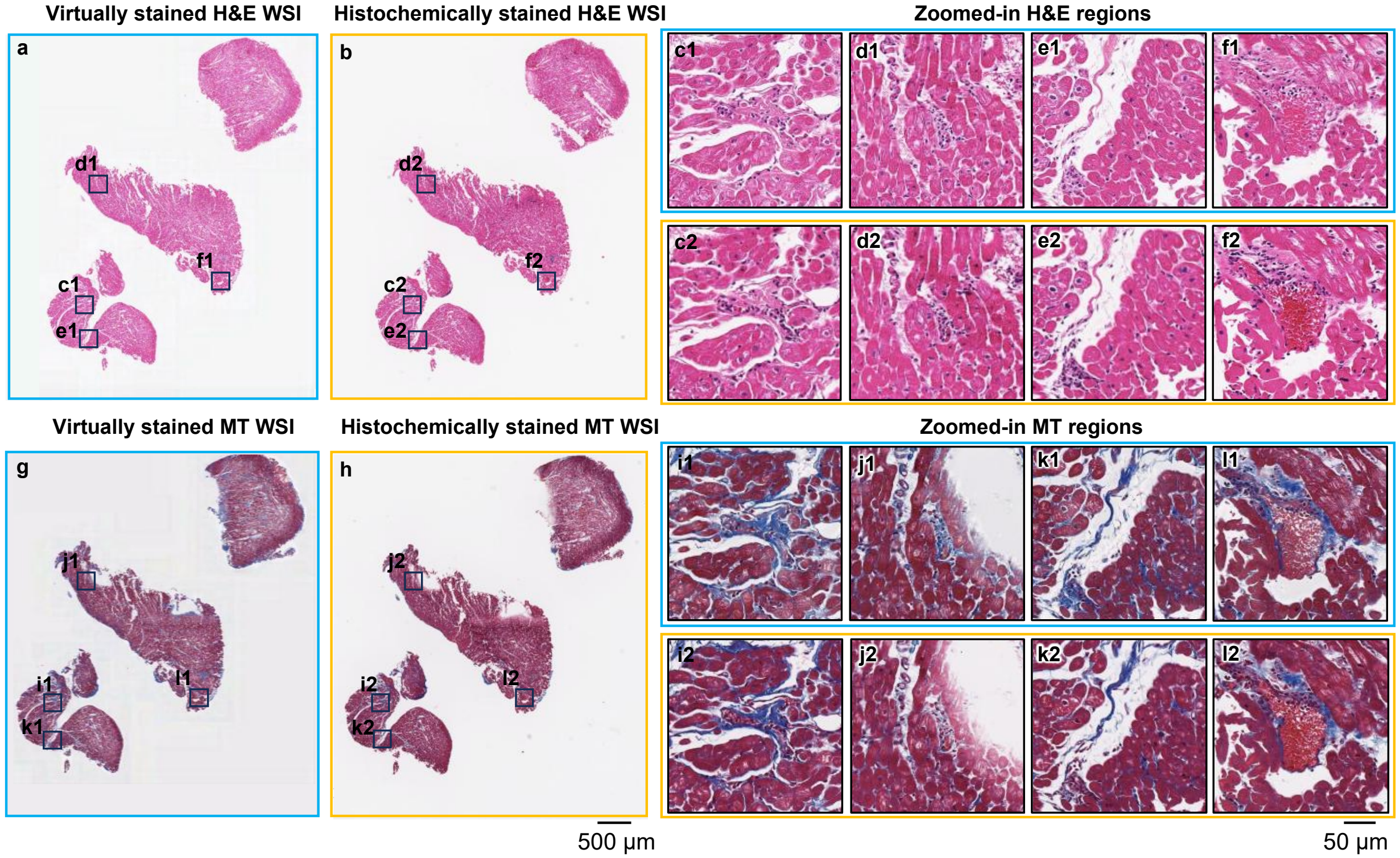

**Figure 3. Visual comparisons between the virtually stained H&E and MT images of a heart transplant rejection sample and their corresponding histochemically stained counterparts.** (a) Virtually stained H&E WSI, which was digitally generated by our virtual staining heart-H&E network by taking label-free autofluorescence images as its input. (b) Histochemical H&E staining results of the same WSI in (a), serving as the ground truth. (c-f) Zoomed-in images of the four exemplary local regions indicated in (a-b), where (c1-f1) are the virtually stained H&E images and (c2-f2) are the corresponding histochemically stained H&E images. (g) Virtually stained MT WSI, which was digitally generated by our virtual staining heart-MT network by taking label-free autofluorescence images as its input. (h) Histochemical MT staining results of the same WSI in (g), serving as the ground truth. (i-l) Zoomed-in images of the four exemplary local regions indicated in (g-h), where (i1-l1) are the virtually stained MT images and (i2-l2) are the corresponding histochemically stained MT images.

## Quantitative evaluations of lung and heart virtual staining results by board-certified pathologists

To further validate the efficacy of our presented deep learning-based virtual staining panels, we conducted score-based, quantitative and blinded studies by board-certified pathologists to assess the image quality of virtual staining. As for lung, 81 virtually stained WSIs (27 H&E WSIs, 27 MT WSIs, and 27 EVG WSIs) generated for 27 unique patients (not seen in the training) and their corresponding 81 standard histochemically stained WSIs were assessed by three board-certified pathologists (WDW, NP, and GX). For the heart tissue virtual staining evaluation, similarly, three board-certified pathologists (WDW, NP, and AJC) assessed 99 virtually stained WSIs, which included 33 H&E WSIs, 33 MT WSIs, and 33 EVG WSIs generated for 33 unique patients (also not seen in the training), together with their corresponding 99 standard histochemically stained WSIs. These virtually stained WSIs and their histochemically stained counterparts were sent to pathologists in two separate batches. The first image batch was given in a mixed fashion (simultaneously containing both virtually stained and histochemically stained WSIs for different cases), and after a three-week washout period, the second image batch generated by the alternative staining method for each case was sent to the diagnosticians. Therefore, if the virtually stained WSI of a case was included in the first batch, its histochemically stained version would be present in the second batch, and vice versa. For each WSI, pathologists were asked to score the image quality without knowing whether it was from a virtual stain or a standard histochemical stain. The scoring metrics for H&E and MT-stained WSIs (for both lung and heart tissue samples) covered three aspects: (1) nuclear detail, (2) cytoplasmic detail, and (3) extracellular detail; and the evaluation criteria for EVG staining (only for lung tissue samples) focused on: (1) elastic tissue highlighting and (2) counterstain quality. For each criterion, pathologists were requested to assign a score to every WSI on a scale from 1 and 4, where 4 indicates "perfect" results, 3 represents "very good", 2 stands for "acceptable", and 1 corresponds to "unacceptable" quality.

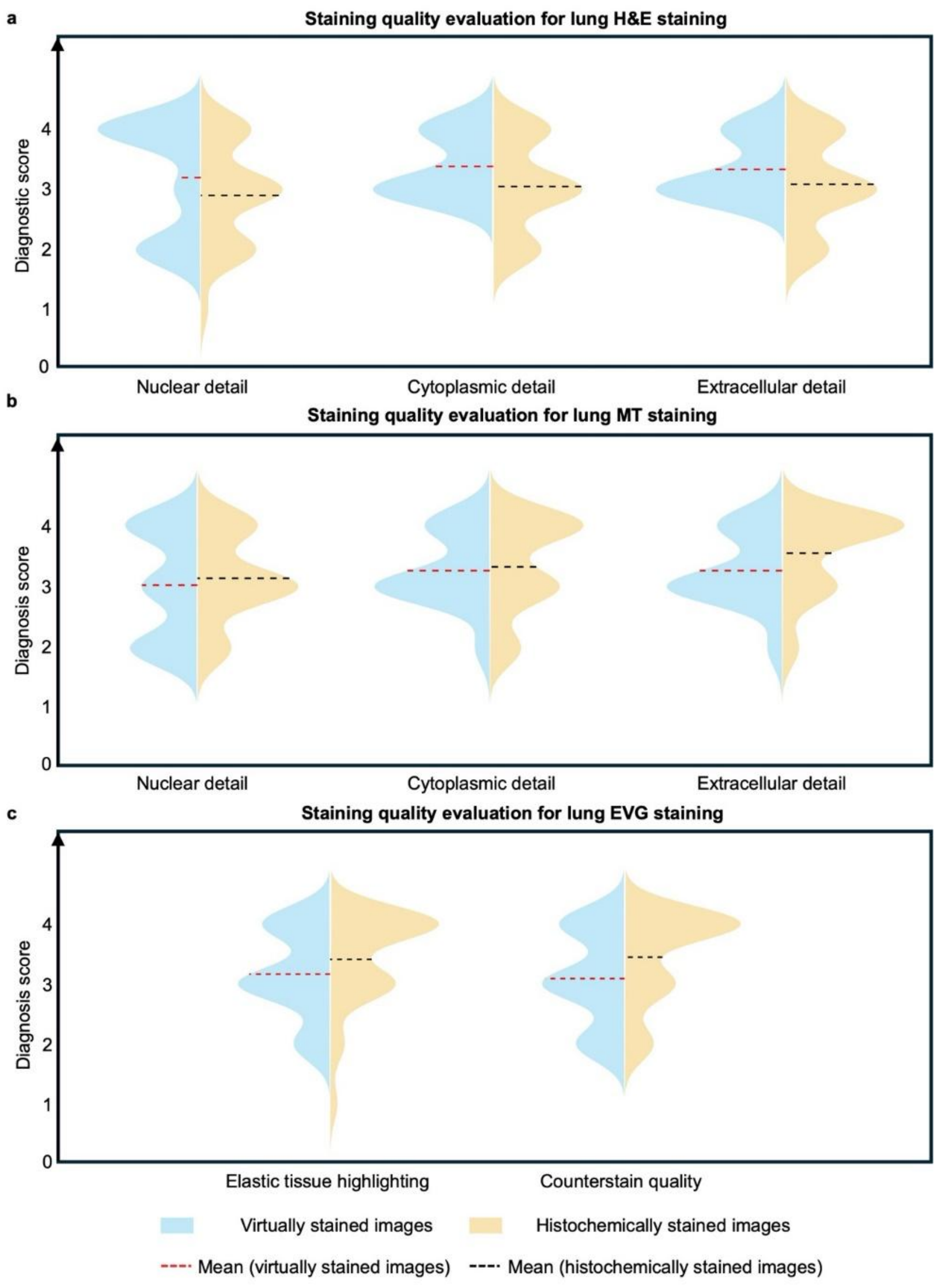

**Figure 4. Score-based quantitative evaluations conducted by three board-certified pathologists for assessing the virtual staining results and their corresponding histochemical ground truth images for lung transplant WSIs. (**a) Violin plots

showing the staining quality scores of virtually and histochemically stained lung tissue H&E images evaluated by three board-certified pathologists. The evaluated aspects include nuclear detail, cytoplasmic detail, and extracellular detail. The staining quality scores range from 1 to 4, where 4 is for perfect, 3 for very good, 2 for acceptable, and 1 for unacceptable. The mean values of these scores for each metric were calculated across all the tested lung cases and all three pathologists (n=27×3=81). (b) Violin plots showing the staining quality scores of virtually and histochemically stained lung MT images evaluated by three board-certified pathologists from the aspects of nuclear detail, cytoplasmic detail, and extracellular detail. The mean values of these scores for each metric were calculated across all the tested lung cases and all three pathologists (n=27×3=81). (c) Violin plots showing the staining quality scores of virtually and histochemically stained lung EVG images evaluated by three board-certified pathologists from the aspects of elastic tissue highlighting and counterstain quality. The mean values of these scores for each metric were calculated across all the tested lung cases and all three pathologists (n=27×3=81).

The quantitative scoring results for the lung virtual staining models, based on the predefined metrics are shown in Fig. 4, where violin plots were employed to compare the distributions of image quality scores between the virtually and histochemically stained WSIs. Specifically, Figs. 4(a), 4(b), and 4(c) illustrate the score distributions for each criterion corresponding to H&E, MT, and EVG stains, respectively. These distributions represent the aggregate scores for each criterion across all the tested lung cases and from all the pathologists. It should be noted that the cases included in this comparative study all featured well-stained histochemical images as high-quality references. Four additional cases with severe histochemical staining artifacts were excluded from our analysis. More details and analyses about these poorly stained, excluded histochemical slides can be found in the Discussion section. Overall, the distributions of the staining quality scores for the virtually stained WSIs and their histochemically stained counterparts show a notable similarity across different aspects, demonstrating the effectiveness of our virtual staining DNN models. It is also noteworthy that the average virtual staining quality scores were slightly higher for the H&E stain, but slightly lower for the MT and EVG stains, compared to their histochemical staining results. To determine whether these differences are statistically significant, we further conducted two-tailed, paired Wilcoxon signed-rank tests[47,48] on the staining quality scores between the virtually and histochemically stained WSIs for different metrics, per pathologist. The results of these statistical analyses are presented in Supplementary Figure 14, along with individual illustrations of the staining quality scores for every metric evaluated by each pathologist. For the H&E stain, the majority (at least two out of three) of the pathologists considered the staining quality of virtually stained WSIs to be significantly superior to their histochemically stained equivalents in terms of nuclear detail, cytoplasmic detail, and extracellular detail. Utilizing the same statistical comparison for the MT and EVG stains, no statistically significant inferiority in the staining quality scores from the virtually stained images, compared to their histochemically stained counterparts, was observed across all the tested/quantified tissue aspects and features. Considering the histochemical cases that were excluded due to poor staining quality, our results suggest that the virtual staining technology is more likely to deliver higher-quality histology images compared to traditional histochemical staining workflows in most histology labs.

Similar to the lung model, the quantitative scoring results for heart tissue samples, based on the established metrics for H&E and MT stains are shown in Fig. 5. Generally, the staining quality scores for the virtually stained WSIs closely parallel those obtained from their histochemically stained counterparts. To ascertain whether the staining quality scores of virtually and histochemically stained high-quality images are statistically significantly different from each other for every stain type, two-tailed, paired Wilcoxon signed-rank tests were conducted on different metrics, per pathologist – following the same analysis reported earlier for the lung tissue samples. The statistical results are presented in Supplementary Figure 15, which also includes detailed illustrations of the staining quality scores for every metric evaluated by each pathologist. Across all the scoring metrics, at least two out of the three pathologists found no statistically significant superiority in the histochemically stained H&E images compared to the virtual H&E staining results, further proving the effectiveness of our heart virtual staining panels. The same conclusion also applies to MT staining, as illustrated in Supplementary Figure 15. Also note that, for the H&E stain, the virtual staining results exhibit statistically significant superior quality in the lung samples ($P$<0.001) and are not significantly inferior in heart samples ($P$=0.081) compared to their histochemically stained counterparts, when averaging different scoring metrics and considering the assessments of all three pathologists by a two-way analysis of variance (ANOVA)[49].

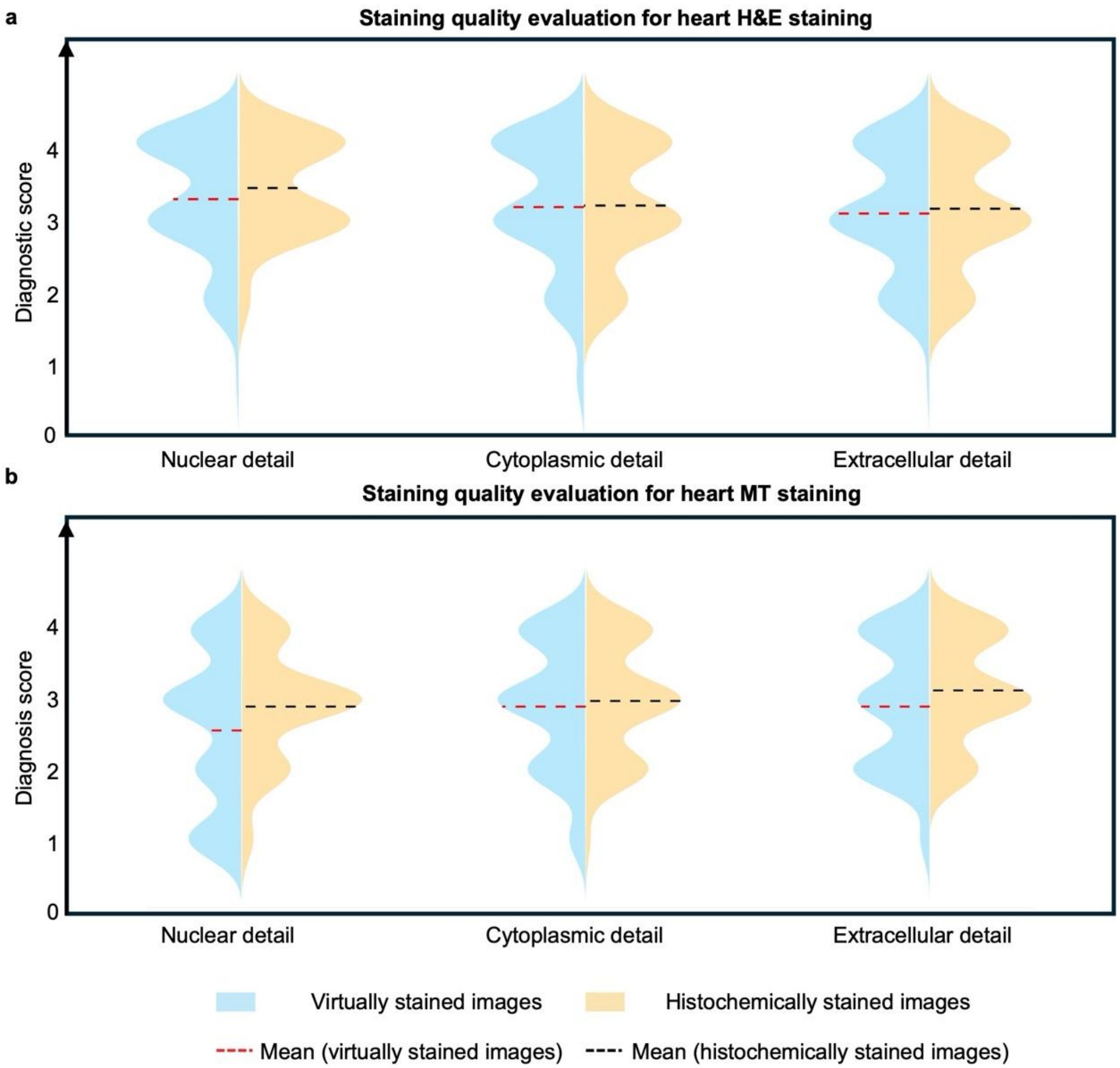

**Figure 5. Score-based quantitative evaluations conducted by three board-certified pathologists for assessing the virtual staining results and their corresponding histochemical ground truth images for heart transplant WSIs. (**a) Violin plots showing the staining quality scores of virtually and histochemically stained heart H&E images evaluated by three board-

certified pathologists. The evaluated aspects include nuclear detail, cytoplasmic detail, and extracellular detail. The evaluated staining quality scores range from 1 to 4, where 4 is for perfect, 3 for very good, 2 for acceptable, and 1 for unacceptable. The mean values of these scores for each metric were calculated across all the tested heart cases and all three pathologists (n=33×3=99). (b) Violin plots showing the staining quality scores of virtually and histochemically stained heart MT images evaluated by three board-certified pathologists from the aspects of nuclear detail, cytoplasmic detail, and extracellular detail. The mean values of these scores for each metric were calculated across all the test heart cases and all three pathologists (n=33×3=99).

**Diagnostic evaluations of virtual stained lung and heart transplant biopsies by board-certified pathologists**

In addition to directly evaluating the staining quality of images created by the virtual staining DNN models, we conducted a blinded study to determine if the virtually stained WSIs could provide diagnostic outcomes (rejection/non-rejection) consistent with those achieved by the conventional histochemically stained WSIs for lung and heart transplant recipients. Before this study, we filtered out 4 cases from the initial set of 27 lung cases and 5 cases from the initial 33 heart cases used for blind testing because they contained less than three tissue fragments, deemed inadequate for a transplant rejection diagnosis as per ISHLT guidelines[1]. For the remaining 23 lung cases and 28 heart cases, three pathologists (WDW, NP, and GX for lung, WDW, NP, and AJC for heart) were provided with the H&E, MT, and EVG WSIs of each lung case (virtual/histochemical), as well as the H&E and MT WSIs of each heart case (virtual/histochemical). They were then asked to classify each case as either positive (indicating rejection) or negative (indicating non-rejection). When all three pathologists reached a unanimous decision (3-0 vote) on specific cases (virtual/histochemical), these diagnostic outcomes were immediately finalized. If consensus was not achieved (2-1 vote), the WSIs (virtual/histochemical) for these cases were forwarded to a fourth, adjudicative pathologist (KAI for lung and KYJ for heart) for additional evaluation. Cases (virtual/histochemical) that received a decisive 3-1 vote among the four pathologists were also finalized, whereas cases (virtual/histochemical) with an even split (2-2 vote) were excluded from the final diagnostic analysis.

The final diagnostic results, which were determined by a decisive voting outcome (either 3-0 or 3-1 votes), are depicted in the confusion matrices shown in Supplementary Figure 16(a) for lung (17 transplant cases in total) and in Supplementary Figure 16(b) for heart (24 transplant cases in total). These matrices compare the diagnostic outcomes from the virtually stained WSIs against those from their corresponding histochemically stained WSIs. A significant proportion of the *diagonal* elements indicate that our virtual staining panels achieved a high concordance in the evaluation of both lung and heart transplant biopsies, matching the diagnostic results of histochemical staining. More specifically, in the case of lung transplant biopsies, the diagnostic results from virtually stained WSIs corresponded with the histochemical staining-based ground truth with a concordance rate of 82.4%. For the heart transplant biopsies, the concordance rate was 91.7%. To address the 6 lung and 4 heart cases that remained tied (2–2 votes), we applied a senior-majority rule: the tie can be broken if the 2 senior pathologists (WDW and KAI for lung; WDW and KYJ for heart) agree on the case. In this situation, the consensus of the 2 senior pathologists of the group of 4 diagnosticians was adopted as the final decision, breaking the tie; in case of a disagreement between the senior pathologists, the tie remained. This approach resolved one true-positive lung case and one true-negative heart case as shown in Supplementary Figure 17(a). After incorporating these resolved tie cases, the overall concordance increased to 83.3% for lung and 92.0% for heart as shown in Supplementary Figure 17(b). Detailed analyses for each case from every pathologist are provided in Supplementary Figure 18 for the lung transplant biopsies and Supplementary Figure 19 for the heart transplant biopsies.

## Discussion

Organ transplantation has evolved from a basic experimental model to an expected choice on the menu of treatments for end-stage organs. For example, lung transplantation has progressed from a rare procedure in the 1980s to a well-accepted option for patients with end-stage lung disease in the modern era[50]. A similar trend was seen in recent years in cardiac transplantation, where the number of transplants has risen to ~5500 procedures per year worldwide with elongated patient survival[51]. Despite the success of heart and lung transplant procedures, the operation itself is only the first part of the patient treatment plan, as there are numerous problems and complications that often develop over the life of a transplant recipient. Among those, transplanted organ rejection is a major cause of morbidity and mortality post-transplantation, with at least a third of lung transplant patients and a quarter of cardiac transplant patients reported to have acute rejections in the first year after transplant[2,52]. Thus, timely detection and diagnosis of rejections are crucial for prompt treatment and improving patient survival rates. Traditionally, diagnosing organ transplant rejection necessitates biopsies followed by multiple histochemical stains to identify histological signs of rejection. This method is usually labor-intensive, reagent-consuming, costly, and requires experienced histotechnologists. Additionally, transplant biopsies are composed of minute tissue samples and new methods to extract more information using less tissue are urgently needed since histochemically stained tissue slides cannot be reused for other ancillary tests. In this study, we developed a virtual staining platform specifically designed for the evaluation of organ transplant biopsies. This platform computationally transforms autofluorescence microscopic images of unlabeled lung and cardiac transplant biopsies into their histologically stained counterparts by multiple DNN models, demonstrating equivalence compared to histochemical staining. We selected autofluorescence microscopy as our input modality because, as one of the most accessible label-free imaging techniques, its utility has been validated in a number of virtual-staining studies[11,53,25,42,32,33]. Moreover, autofluorescence imaging modules are already incorporated into commercial whole slide scanners[54–56] used in digital pathology workflow, obviating the need for extra optical components or increased costs, also enabling seamless integration with the current digital pathology systems. The inclusion of all four autofluorescence channels—DAPI, TxRed, FITC, and Cy5—using standard filter sets is important to achieving a good performance in our virtual staining models. Omitting a channel—or multiple channels—will result in reduced image fidelity, as prior studies[42,57,58] have already shown that fewer autofluorescence channels compromise the model's ability to consistently reveal histological features. Our models were capable of virtually staining lung transplant samples using 3 stains: H&E, MT, and EVG. Additionally, they could virtually stain cardiac samples using H&E and MT. H&E is the main histochemical stain utilized in transplant pathology and serves as the initial stain for the pathologist's microscopic examination, facilitating the observation of tissue architecture and quantification of the extent and location of inflammatory cells. MT is employed to assess the degree of fibrosis within the transplanted organ, while EVG highlights residual elastic tissue, aiding in the differentiation between scar tissue and a small, occluded blood vessel. Our label-free platform can routinely generate high-quality, virtually stained WSIs for multiple stains, ready for review with a rapid speed of $<30$ seconds per $\text{mm}^2$, which can be further improved with a more advanced digital computing environment. Our virtually stained WSIs demonstrated high concordance and exhibited no statistically significant differences compared to their histochemically stained counterparts for most organ-stain combinations and demonstrated statistically significant superiority in lung H&E samples, as validated through two-tailed Wilcoxon signed-rank tests for different staining quality metrics. In fact, in favor of the histochemical ground truth stains, slides with severe histochemical staining artifacts were excluded from our analyses, and therefore we only compared our virtual staining results against high-quality slides. It is also important to note that, the success of virtual staining models for the accurate evaluation of lung and heart transplant biopsies fundamentally rely on the precise approximation of each individual pixel with respect to the histochemical ground truth images, comprising over 550 billion pixels of tissue images in total – which constitute the labels of our test set.

Our results and analyses underscore that virtual staining of label-free tissue can deliver microscopic images with diagnostic content comparable to those of the traditional histochemical staining workflow. A concordance rate of ~80–95% between diagnoses from virtually stained images and their histochemical counterparts already presents a robust demonstration of diagnostic equivalence, considering that transplant-rejection biopsies are subject to unavoidable inter- and intra-pathologist variability[59–61], and section-to-section heterogeneity. Furthermore, our models allow for the computational transformation of a single label-free tissue slide into multiple histological stains, rapidly providing a panel of microscopic images that can be used for diagnosis. As a result, our platform not only preserves precious biopsy tissue for further molecular testing and genomic assays, but also markedly decreases the time required for additional tissue cuts and MT/EVG staining procedures; the waiting time for these special stains, following the H&E-based initial examination, can significantly extend the overall turnaround time in clinical settings. Moreover, in the case of multiplexed virtual staining, the generated virtually stained WSIs are perfectly registered with each other at the nanoscale. This is because they originate from the same label-free tissue slide. This is highly beneficial for pathologist evaluations because they will not encounter tissue heterogeneity-related structural mismatches between adjacent slides stained with different stains, which is common in histochemical staining.

Another significant advantage of virtual staining models lies in their stability and repeatability, which enable consistent delivery of high-quality histological images; in contrast, traditional histochemical staining sometimes encounters technical artifacts. This is specifically important for the identification and grading of lung and heart rejection cases based on tiny tissue fragments, which can easily be crushed and distorted during tissue preparation and histochemical staining processes. This can lead to various artifacts, such as tissue compression and hemorrhage, and suboptimal staining quality. These non-ideal morphological changes can significantly alter tissue morphology, complicate histological evaluations, and mimic rejection histology, thereby resulting in diagnostic difficulties. In this context, virtual staining of label-free tissue can minimize these artifacts, enabling pathologists to arrive at a definitive diagnosis. For example, four histochemically stained slides from one of the histology laboratories used in our study exhibited prominent suboptimal staining quality for the H&E stain. These slides featured extensive pale, under-stained areas and lacked well-defined nuclear contours, as shown in Supplementary Figure 20 (a, bottom two rows). In contrast, their virtually stained equivalents generated by our DNN model demonstrated significant improvements in image quality. This is evident in the enhanced contrast between the nuclei and surrounding cytoplasmic-extracellular features, as shown in Supplementary Figure 20(a, top two rows). The superior image quality of virtual staining was further supported by the pathologist scoring results. Supplementary Figure 20(b) compares different staining quality metrics for the virtually and histochemically stained images. It showcases these four lung cases with notable histochemical staining failures, where a significant score superiority of virtual staining was observed. Moreover, even though these cases were excluded from the previous score-based evaluations of the staining quality reported in the Results section (Fig. 4), the virtually stained H&E images were still found to have higher quality scores (statistically significant) compared to their histochemically stained counterparts, with most pathologists favoring virtual staining over histochemical staining for the H&E stain applied for lung biopsies. Although similar histochemical staining artifacts were not observed in our cohort of heart transplant samples, we believe that virtual staining generally mitigates staining issues related to tiny heart tissue fragments as well. This advantage stems from our rigorous training image dataset curation, which minimizes artifacts and prevents contamination at every stage, allowing virtual staining to outperform conventional histochemical staining methods in certain cases. First, the high-quality WSIs used for training were preselected and picked by a senior board-certified pathologist. Then, to generate input-target training pairs, we applied iterative training-registration cycles (see Methods for details) to filter any image pairs exhibiting low similarity (with a peak signal-to-noise ratio (PSNR) < 16 or the structural similarity index (SSIM[62]) < 0.6) after conversion to the same brightfield domain. This step effectively excludes samples with tissue artifacts, such as tears, cracks, tissue folding, or out-of-focus areas on the input or target domains. Finally, the remaining pairs were manually reviewed by the authors to ensure a clean, artifact-free image dataset. These stringent quality-control operations yield high-quality training pairs that underlie the superior performance of our virtual-staining models. Additionally, using perfectly paired images of the same tissue section before and after histochemical staining—rather than serial sections—combined with pixel-wise loss constraints served as another key factor in ensuring the high fidelity of our virtual staining models.

Additionally, virtual staining is able to generate histologically stained images with a more uniform or standardized color distribution. Histochemical staining often exhibits inevitable color variations across different clinical laboratories, and even on slides stained in different periods within the same lab. For example, Supplementary Figures 21(a) and 21(b) present several virtually stained H&E images with uniform colorization alongside their corresponding histochemically stained counterparts, which display diverse colorization styles, for both lung and heart tissue samples, respectively. A similar consistency in color style was also observed with the MT stain, as shown in Supplementary Figure 22. This style uniformity in virtual staining can mitigate the inherent variability of histological staining processes, potentially enhancing the diagnostic workflow by providing clinicians with consistently stained slides.

The abovementioned advantages of virtual staining over histochemical staining, such as reduced staining artifacts and decreased stain variability, hold significant potential for advancing the development of new diagnostic algorithms for the evaluation of organ transplant rejection using biopsies. As the number of organ transplantations is steadily increasing each year, along with the high prevalence of transplanted organ rejection, there is a strong interest among researchers, health specialists and pharmaceutical companies in exploring novel ways for early detection and precision treatments to prevent rejection. Given that histological analysis is the gold standard for diagnosis, many of these research initiatives employ computational image analysis tools and deep learning models based on histochemically stained images to detect subtle clues for the organ rejection process. However, the extraction of small tissue fragments can induce morphological artifacts that along with suboptimal processing and staining quality, can render parts of the samples unsuitable for computational image exploration. This could even fail the learning of artificial intelligence-based image analysis models by misleading them to recognize various induced artifacts instead of the subtle morphological clues present in biopsies. Additionally, the inherent variability in histochemical staining can further complicate the training of such detection models. In contrast, virtual staining mitigates these issues by providing histology images in a uniform and standardized style, which facilitates model convergence without the need for extensive preprocessing. Therefore, virtual staining can greatly benefit these downstream organ rejection evaluation algorithms.

In conclusion, we developed deep learning-based virtual staining panels specifically designed for the evaluation of organ transplant biopsies and demonstrated the feasibility of virtual staining of minute tissue samples, as seen in heart and lung transplant biopsies. Through computational analysis and evaluation by pathologists, we demonstrated that our model could generate virtually stained WSIs in minutes, with high repeatability and diagnostic concordance compared to traditional histochemical staining workflow. However, as noted, rejection (positive) cases in our lung and heart cohorts are scarce, and most samples are negative due to data limitations. Future research will prioritize acquiring additional positive rejection samples to achieve a more balanced cohort and will incorporate diverse samples, such as transplant biopsies from different stages of rejection, in order to showcase the utility of virtual staining models for accurate detection of organ transplant rejection across multiple grades.

# Materials and Methods

**Sample preparation and standard histochemical staining of H&E, MT and EVG**

The lung and heart transplant tissue samples used in this study were sourced from pre-existing, de-identified tissue blocks at UCLA Translational Pathology Core Laboratory (TPCL), authorized under IRB approval number 18-001029. The study included lung specimens from 45 distinct patients and heart samples from 51 distinct patients. For each patient, three adjacent lung slices (and two for the heart), each approximately 4 μm thick, were serially sectioned from unlabeled tissue blocks, deparaffinized, and mounted on glass slides. Following autofluorescence imaging, these unlabeled adjacent tissue sections were sent to TPCL and the Department of Anatomic Pathology at Cedars-Sinai Medical Center (Los Angeles, CA, USA) for standard histochemical H&E and MT staining, and to the Keck School of Medicine of the University of Southern California for histochemical EVG staining.

**Image data acquisition**

Autofluorescence images of the unlabeled lung and heart transplant tissue slides were captured using a Leica DMI8 microscope equipped with a 40×/0.95 NA objective lens (Leica HC PL APO 40×/0.95 DRY) under the control of Leica LAS X microscopy automation software. The label-free microscopic imaging employed four fluorescence filter cubes: DAPI (Semrock OSFI3-DAPI-5060C, EX377/50 nm, EM 447/60 nm), TxRed (Semrock OSFI3-TXRED-4040C, EX

562/40 nm, EM 624/40 nm), FITC (Semrock FITC-2024B-OFX, EX 485/20 nm, EM 522/24 nm), and Cy5 (Semrock CY5-4040C-OFX, EX 628/40 nm, EM 692/40 nm). Each image was acquired using a scientific complementary metal-oxide-semiconductor (sCMOS) image sensor (Leica DFC 9000 GTC) with an exposure time of ~300 ms for all four autofluorescence channels. After undergoing standard histochemical H&E, MT, and EVG staining procedures, the tissue slides were digitized using a brightfield digital pathology slide scanner (Leica Biosystems Aperio AT2).

**Dataset division and preparation**

From our cohort of 45 unique, de-identified lung transplant recipients, tissue slides from 18 patients were designated for training, with the remainder from 27 patients reserved for blind testing. This selection was guided by a senior board-certified pathologist (W.D.W), ensuring an even distribution of rejection and non-rejection statuses among the training samples—half rejection and the other half non-rejection. Similarly, for the heart models, slides from 18 unique heart transplant recipients were allocated for training and slides from 33 additional patients were set aside for blind testing. However, due to a scarcity of rejection samples in our heart transplant sample cohort, all 18 heart training slides were non-rejection cases. After the data division, the paired autofluorescence WSIs and their corresponding brightfield histochemically stained WSIs underwent a meticulous three-step registration workflow to produce precisely aligned local FOV pairs for training. Initially, a rigid registration at the WSI level was performed by calculating the maximum cross-correlation coefficient between each autofluorescence WSI and its corresponding histology brightfield WSI. This computation estimated the relative rotation angle and shift distance between the WSIs, which were used to spatially transform the histochemically stained WSI for better alignment with its autofluorescence counterpart. In the second phase, a finer registration process was undertaken at the image patch level. This involved dividing the coarsely aligned WSI pairs into smaller FOV pairs of dimensions 3248×3248 pixels (~528×528 μm²). A multi-modal affine image registration[63] was then executed to correct shifts, size alterations, and rotations between the histology image FOVs and their autofluorescence equivalents. Moreover, before cropping the WSIs into smaller local FOVs, intensity normalization was applied to each autofluorescence channel, which was achieved by dividing the original WSI by a background image obtained through 2-D Gaussian filtering with a kernel that had a standard deviation of 300 pixels. Subsequently, the registered histological image FOVs (each 3248×3248 pixels) were center-cropped to 2048×2048 pixels (~333×333 μm²) to eliminate potential edge artifacts. Then, to achieve pixel-level alignment, the third phase involved an iterative registration step based on an elastic pyramid cross-correlation algorithm[11,64,65]. This step required an initial virtual staining network trained and applied to the autofluorescence images. The resulting roughly stained images were then compared with their bright-field histochemically stained counterparts using the elastic pyramid registration algorithm, resulting in transformation maps which were applied to correct local discrepancies in the ground truth images to better align with their corresponding autofluorescence images. Then, these better registered image pairs were utilized to train a new iteration of the rough virtual staining network, followed by a new round of elastic registration. This training-registration cycle was repeated 3-5 times until the autofluorescence inputs and the bright-field ground truth image patches were precisely matched at the single-pixel level. A manual data cleaning process was subsequently conducted to eliminate image pairs with artifacts such as tissue tearing (during standard histochemical staining) or defocusing (during digital imaging). Finally, our three virtual staining DNN models for lung were trained on datasets comprising 1051, 1041, and 851 2048×2048-pixel (~333×333 μm²) paired microscopic image patches for H&E, MT, and EVG stains, respectively. For the heart, the H&E and MT virtual staining DNN models were trained using 1240 and 1415 similar-sized paired image patches, respectively. Furthermore, for the heart virtual staining models, the histochemically stained ground truths used in training were sourced from two different staining institutions, resulting in variations in H&E and MT color styles. To standardize these styles before feeding them into the training cycles of the virtual staining networks, we developed separate style transfer neural networks for H&E and MT stains based on the CycleGAN[26,66] training framework. These networks were trained to adapt the staining style from the first institution (Cedars-Sinai Medical Center) to match that of the second (UCLA TPCL). The images with the adjusted second style, along with those in the original style from the second institution, were then utilized as ground truths for training the virtual staining networks; see Supplementary Figure 23. Throughout each epoch of the training process, these paired image FOVs were further subdivided into smaller patches of 256×256 pixels, normalized to have zero mean and unit variance, and augmented through random flipping and rotation operations.

**Network architecture, loss function and training schedule**

To develop the virtual staining DNN models for lung and heart transplant tissue samples, we employed a structurally-conditioned GAN architecture. This architecture consists of two components: a virtual staining generator network G and a discriminator network D, as detailed in Supplementary Figure 1(a) and 1(b), respectively. The generator network adopts a five-level U-net[67] structure, designed to learn the statistical transformation from autofluorescence input images to their corresponding brightfield histochemically stained images; and the discriminator network is a convolutional neural network classifier aimed to differentiate between the virtually stained images produced by the generator and the actual histochemically stained images. This competitive training process fosters alternating optimization of the generator and discriminator, reaching a stable equilibrium where both networks are effectively trained.

The loss used for training the generator network G was defined as:

$$L_{\mathrm{G}} = L_1\{I_{\mathrm{VS}}, I_{\mathrm{HS}}\} + \alpha \cdot TV\{I_{\mathrm{VS}}\} + \beta \cdot \left(1 - D(I_{\mathrm{VS}})\right)^2 \quad (1).$$

where, $I_{\mathrm{VS}}$ and $I_{\mathrm{HS}}$ correspond to the virtually stained image output by the generator network G and its corresponding histochemically stained equivalent, respectively. $D(\cdot)$ represents the probability of being a histochemically stained image predicted by the discriminator network. The coefficients $\alpha$ and $\beta$ were empirically set to 0.02 and 100, respectively. The $L_1\{\cdot\}$ stands for the mean absolute error loss, defined as:

$$L_1\{I_{\mathrm{VS}}, I_{\mathrm{HS}}\} = \frac{1}{M \cdot N} \sum_{m,n} |A(m,n) - B(m,n)| \quad (2),$$

where $M$ and $N$ represent the pixel numbers at the horizontal and vertical directions of the images $A$ and $B$, i.e., $I_{\mathrm{VS}}$ and $I_{\mathrm{HS}}$ in our implementation, while $m$ and $n$ are pixel indices.

$TV\{\cdot\}$ stands for the total variation (TV) loss, which can be written as:

$$TV\{A\} = \sum_{m,n} (|A(m+1,n) - A(m,n)| + |A(m,n+1) - A(m,n)|) \quad (3),$$

where $m$ and $n$ are the coordinates of the image $A$, i.e., $I_{\mathrm{VS}}$ in our implementation.

The loss used for training the discriminator network (D) was defined as:

$$L_{\mathrm{D}} = D(I_{\mathrm{VS}})^2 + \left(1 - D(I_{\mathrm{HS}})\right)^2 \quad (4).$$

The generator network G and the discriminator network D were updated at a frequency ratio of 3:1. The learning rate was set as $10^{-4}$ for optimizing network G, while the learning rate for optimizing network D was set as $10^{-5}$. The Adam optimizer[68] was used for the network training. The batch size was set as 8. Upon convergence of the networks, the optimal model of network G was selected based on the lowest $L_{\mathrm{G}}$ value, which was further corroborated by a human-based visual evaluation of the validation image set.

**Statistical analysis**

A two-way ANOVA[49] was utilized to determine if there were significant differences between the virtually stained WSIs and their histochemically stained counterparts by taking the average scores of different staining quality metrics and incorporating inputs from all three pathologists. Additionally, for each pathologist, a two-tailed paired Wilcoxon signed-rank test was performed to determine the statistical significance of the differences between the virtual and histochemical staining results for each specific staining quality metric. It is noted that for all the tests, a *P* value of $<0.05$ was considered to indicate a statistically significant difference. All the analyses were performed using IBM SPSS Statistics v29.0.

**Other implementation details**

For the pathologist-involved blind quantitative evaluation, as reported in Figs. 4-5 and Supplementary Figures 14-20, an online image-sharing platform[69] was utilized to enable pathologists to view and assess the virtual and histochemical WSIs of lung and heart transplant cases. All the image preprocessing and registration operations were performed using MATLAB, version R2022b (MathWorks). The codes for training the virtual staining DNN models were developed in Python 3.8.15, utilizing TensorFlow 2.5.0. All the network training and testing tasks were carried out on a desktop computer equipped with an Intel Core i9-10920X CPU, 256GB of memory, and an Nvidia GeForce RTX 3090 GPU.